\pdfoutput=1
\documentclass[a4paper,11pt]{iopart}
\usepackage{graphicx}
\usepackage[utf8]{inputenc}		
\usepackage[T1]{fontenc}               
\usepackage{cite}
\usepackage{color}
\usepackage{amsfonts}
\usepackage{wasysym}
\usepackage{srcltx}
\usepackage{hyperref}
\begin{document}
\title{Playing Music in Just Intonation -- \\A Dynamically Adapting Tuning Scheme}
\author{Karolin Stange$^{1,2}$, Christoph Wick$^3$, and Haye Hinrichsen$^1$}
\address{$^1$ Fakult\"at f\"ur Physik und Astronomie, Universit\"at W\"urzburg\\ Am Hubland, 97074 W\"urzburg, Germany}
\address{$^2$ Hochschule f\"ur Musik, Hofstallstr. 6–8, 97070 W\"urzburg, Germany}
\address{$^3$ Fakult\"at f\"ur Mathematik und Informatik, Universit\"at W\"urzburg\\ Am Hubland, 97074 W\"urzburg, Germany}
\ead{hinrichsen@physik.uni-wuerzburg.de}

\begin{abstract}
We investigate a dynamically adapting tuning scheme for microtonal tuning of musical instruments, allowing the performer to play music in just intonation in any key. Unlike other methods, which are based on a procedural analysis of the chordal structure, the tuning scheme continually solves a system of linear equations without making explicit decisions. In complex situations, where not all intervals of a chord can be tuned according to just frequency ratios, the method automatically yields a tempered compromise. We outline the implementation of the algorithm in an open-source software project that we have provided in order to demonstrate the feasibility of the tuning method.

\end{abstract}

\def\d{{\rm d}}
\def\0{\emptyset}

\def\alphavec{{\vect{\alpha}}}

\def\comment#1{\color{red}[\textbf{comment: #1}]\color{black}}
\def\mark#1{\color{red}#1 \color{black}}

\def\ET{{\rm\scriptscriptstyle ET}}
\def\JI{{\rm\scriptscriptstyle JI}}
\def\OPT{{\rm\scriptscriptstyle opt}}
\def\REF{{\rm\scriptscriptstyle ref}}

\pagestyle{plain}

\section{Introduction}

\begin{figure}[t]
\centering\includegraphics[width=150mm]{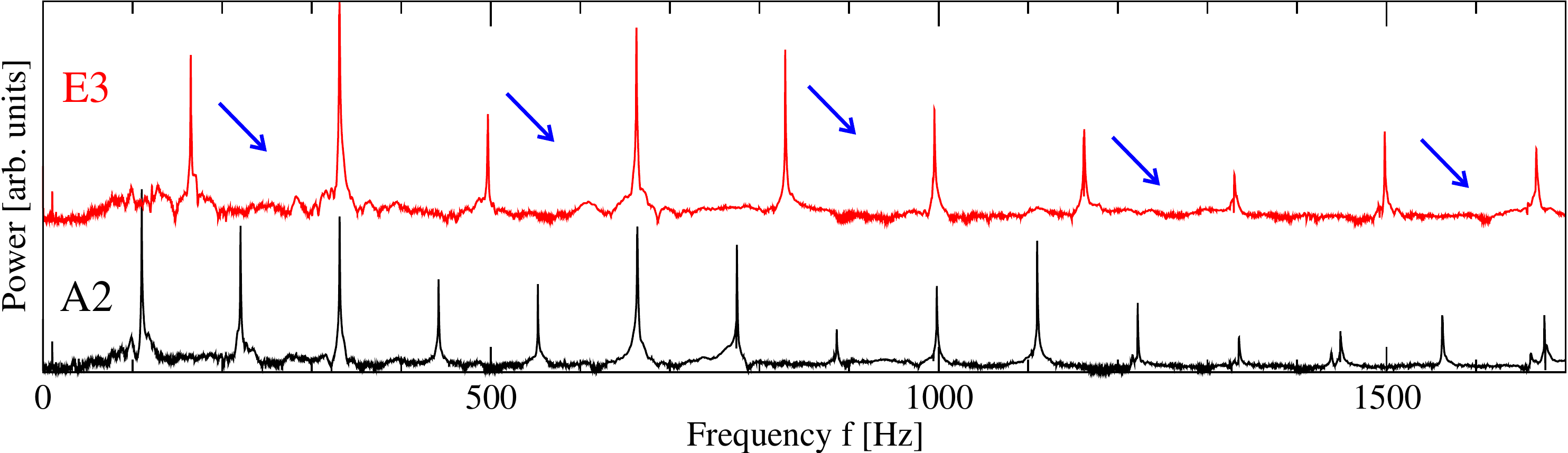}
\caption{Consonance of a just perfect fifth. The figure shows a measured power spectrum of the piano keys A2 (110 Hz) and E3 (165 Hz). As first theorized by Helmholtz (1877)~\cite{helmholtz}, the fifth is perceived as consonant because many partials of the corresponding natural harmonic series coincide (marked by the arrows in the figure).\label{spectra}}
\end{figure}

The first attempts to mathematically characterize musical intervals date back to Pythagoras, who noted that the consonance of two tones played on a monochord can be related to simple fractions of the corresponding string lengths (for a general introduction see e.g. \cite{Geller,White}). Physically this phenomenon is caused by the circumstance that oscillators such as strings do not only emit their fundamental ground frequency but also a whole series of partials (overtones) with integer multiples of the ground frequency. Consonance is related to the accordance of higher partials, i.e., two tones with fundamental frequencies $f$ and~$f'$ tend to be perceived as consonant if the $m$-th partial of the first one matches with the $n$-th partial of the second, meaning that $mf=nf'$ (see Fig. \ref{spectra}). Although the perception consonance is a highly complex psychoacoustic phenomenon (see e.g.~\cite{mcdermott,stolzenburg}) which also depends on the specific context~\cite{milne,parncutt}, one can basically assume that the impression of consonance is particularly pronounced if $m$ and $n$ are small. Examples include the perfect octave $(m/n=2/1)$, the perfect fifth ($3/2$), and the perfect fourth ($4/3$). Larger values of $m$ and $n$ tend to correspond to more dissonant intervals. If a normally consonant interval is sufficiently detuned from the ``just'' tuning, i.e., the simple frequency ratio, the resulting mismatch of almost-coinciding partials leads to a superposition of waves with slightly different frequencies~\cite{helmholtz}. The fast beating of these almost-coinciding partials can result in a sensation of roughness or being out of tune that ruins the perception of consonance.

With the historical development of fretted instruments and keyboards, it made sense to define a system of fixed frequencies in an octave-repeating pattern. The frequency ratios of stacked intervals multiply. (For example, the chord A2-E3-B3, consisting of two perfect fifths each having the ratio $3/2$, yields a frequency ratio of $9/4$ from A2 to B3.) This immediately confronts us with the fundamental mathematical problem that multiplication and prime numbers are incommensurate in the sense that powers of prime numbers never yield other simple prime numbers. For example, it is impossible to match $k$ just perfect fifths with $\ell$ just octaves since $(3/2)^k \neq (2/1)^\ell$ for all $k,\ell\in \mathbb{N}$. Mathematically speaking, the concatenation of just musical intervals (by multiplying their frequency ratios) is an operation that does not close up on any finite set of tones per octave.

Fortunately the circle of fifths approaches a closure up to a small mismatch: when stacking twelve just fifths on top of each other the resulting frequency ratio $(3/2)^{12}$ differs from seven octaves (ratio $2^7$) only by a small amount; explaining why the Western chromatic scale is based on twelve semitones per octave. Nevertheless the remaining difference of $\approx 1.4\%$ ($23.46$ cents\footnote{In music theory, a cent ($\cent$) is defined as 1/100 of a semitone in the equal temperament.}), known as the Pythagorean comma, is clearly audible and cannot be neglected in a scale with fixed frequencies.  Likewise, a sequence of four just fifths transposed back by two octaves ($(3/2)^4/2^2$) differs from a major third of $5/4$ by the so-called syntonic comma of $21.51$ cents.

Since it is impossible to construct a musical scale that is based exclusively on pure beatless intervals, one has to seek out suitable compromises. Over the centuries this has led to a fascinating variety of tuning systems, called temperaments, which reflect the harmonic texture of the music in the respective epoch~\cite{Barbour}. With the increasing demand of flexibility the equal temperament (ET) finally prevailed in the 19$^{\rm th}$ century and has established itself as a standard temperament of Western music. In the ET the octave is divided into twelve equally sized semitones of constant frequency ratio $2^{1/12}$. The homogeneous geometric structure of the ET ensures that all interval sizes are invariant under transposition (displacements on the keyboard). This means that music can be played in any key, differing only in the global pitch but not in the harmonic texture. 

However, this high degree of symmetry can only be established at the expense of harmony~\cite{DuffinBook}. In fact, the only just interval in the ET is the octave with the frequency ratio $2/1$ while all other intervals are characterized by irrational frequency ratios, deviating from the just intervals. For some intervals the variation is quite small and hardly audible, e.g., the equally tempered fifth differs from the just perfect fifth by only two cents. For other intervals, however, the deviations are clearly audible if not even disturbing. For example, the minor third in the ET is almost $16\cent$ smaller than the natural frequency ratio 6:5. The same applies to the major third which is about $14\cent$ greater than the ratio 5:4. These discrepancies may explain why there was some reluctance among many musicians to accept the ET. It was not until the 19th century that the ET became a new tuning standard, presumably both because of Western music's increasingly complex harmonies and because of an increasing intonational tolerance on the part of the audience.\\

\noindent\textbf{Just Intonation}
~\newline~\newline
Although musical temperaments provided a good solution for most purposes, music theorists and instrument makers have searched for centuries for possible ways to overcome the shortcomings of temperaments, aiming to play music solely on the basis of pure intervals, referred to as \textit{just intonation} (JI)~\cite{DuffinArticle}. Tuning an instrument in just intonation means to adjust the twelve pitches of the octave in such a way that all frequency ratios with respect to a certain reference frequency $f^*$ are given by simple rational numbers. For example, a possible choice of such frequency ratios is listed in Table~\ref{tab:frequencyratios}.

\begin{table}
\begin{center}
\begin{tabular}{l|cccccccccccc}
semitones& 0 & 1 & 2 & 3 & 4 & 5 & 6 & 7 & 8 & 9 & 10 & 11\\ \hline
$f/f^*$ & 1 & 16/15 & 9/8 & 6/5 & 5/4 & 4/3 & 45/32 & 3/2 & 8/5 & 5/3 & 9/5 & 15/8 
\end{tabular}
\end{center}
\caption{The most common choice of frequency ratios in just intonation, known 5-limit tuning. }
\label{tab:frequencyratios}
\end{table}

Just intonation always refers to the tonic of a given scale, referred to in this article as the \textit{keynote}. In its own reference scale, JI sounds very consonant if not even sterile, but unfortunately a transposition in other scales is not possible. For example, tuning a piano in just intonation with keynote C,  a C-major triad sounds consonant while most triads in other keys appear to be out of tune. The same applies to modulations from one key to another. Thus, for good reasons, just intonation has the reputation of being absolutely impractical.

To overcome this problem, a possible solution would be to enlarge the number of tones per octave. Important historical examples are for example a keyboard with 19 keys per octave suggested by the renaissance music theorist G. Zarlino (1558), and the two-manual archichembalo by N. Vicentino (1555) with 36 keys per octave~\cite{Vicentino}. More recent examples include the Bosanquet organ with 48 keys per octave (see~\cite{helmholtz}), the harmonium with 72 keys per ocatave by~\cite{Oettingen}, and the 31-tone organ by~\cite{fokker}. Today various types of electronic microtonal interfaces are available~(see \cite{Douglas,macritchie}). However, as one can imagine, such instruments are very difficult to play.

\vspace{2mm}
Temperaments are primarily relevant for keyboard instruments (such as piano, harp, organ) and fretted instruments (e.g. lute and guitar), where all tones are tuned statically in advance. In comparison many other instruments (e.g. string instruments) allow the musician to recalibrate the pitches during the performance, and the same applies of course to the human voice. Musicians playing such instruments tune the pitches \textit{dynamically} while the music is being played. By listening to the harmonic consonance and its progression, well-trained musicians are able to estimate the appropriate frequency intuitively and to correct their own pitch instantaneously. Usually the played notes are a compromise between just intervals and the prevailing ET~\cite{DuffinArticle}. By dynamically adapting the pitches, this allows one to significantly improve the harmonic texture. In this respect the following quotation of the famous cellist Pablo Casals~\cite{Casals} seems to make a lot of sense: 
\begin{quote}
\textit{``Don’t be scared if your intonation differs from that of the piano. It is the piano that  is out of tune.  The piano with its tempered scale is a compromise in intonation.''} 
\end{quote}
\vspace{2mm}

\noindent\textbf{Dynamically Adapting Tuning Schemes}
~\newline~\newline
Is it possible to mimic the process of dynamical tuning by constructing a device which instantaneously calculates and corrects the pitches like a singer in a choir? This idea can be traced back to the early days of electronic keyboard instruments. Since then various implementations have been suggested, the most important ones including \textit{GrovenMax}~\cite{GrovenMax}, \textit{Justonic Pitch Palette}\texttrademark~\cite{Justonic}, \textit{Mutabor}~\cite{Mutabor}, \textit{Hermode Tuning}\texttrademark~\cite{Hermode}, and \textit{TransFormSynth}~\cite{Sethares}. These will be described in the following.

\begin{itemize}
\item One of the first pioneers of dynamic tuning schemes was the Norwegian microtonal composer and music-theorist Eivind Groven~\cite{GrovenMax}. In 1936 he constructed a pipe organ driven by an electric circuit of relays used in telephone switchboards at that time.The organ had three sets of pipes differing by a syntonic comma. Playing a chord on the manual, the electro-mechanical logical circuit computed the optimal arrangement of the chord and triggered the pipes accordingly. In 1995 this method was implemented on a computer, redirecting the output of a MIDI keyboard to three MIDI pianos differing in pitch by a syntonic comma~\cite{Premiere}.

\item\textit{Justonic Pitch Palette}\texttrademark ~was proprietary software produced by Justonic Tuning, Inc. based on a patented method developed by J. William Gannon and Rex A. Weyler~\cite{Justonic}. This tuning method is dynamic in so far as the artist himself can change the keynote frequency $f^*$ during the performance by hand. The corresponding frequencies are then retrieved from a table and sent to a microtonal synthesizer. The selection of the keynote requires additional hardware such as an extra manual.

\item\textit{Mutabor} is a microtonal software project initiated by M. Vogel at the University of Darmstadt~\cite{Mutabor}. The first version, referred to as Mutabor I, was designed as ``\textit{a system with 171 steps per octave for electronic keyboard instruments}''. Depending on the actual chord being played, the program calculates the actual chordal root and tries to tune all frequencies in pure fifths, fourths, and thirds. However, this leads to audible frequency shifts between subsequent chords. From 1987 on Mutabor II improved this method, aiming to produce a usable PC-software for MIDI devices and allowing the user to develop individual tuning algorithms. In a third stage starting in 2006, Mutabor has evolved into a full-fledged microtonal programming language. Nevertheless, for music with a greater harmonic complexity, the chordal root is not always detected reliably. As a solution, Mutabor offers the user to pre-determine the succession of keynotes in a separate MIDI file.

\item \textit{Hermode Tuning}\texttrademark ~is a commercial adaptive tuning scheme developed by Werner Mohrlok~\cite{Hermode}. To our knowledge it is the only adaptive tuning scheme that has reached a wider dissemination, ranging from implementations in church organs to plugins for software packages such as \textit{Cubase}\texttrademark. Instead of determining the chordal root, the algorithm tunes intervals between adjacent tones of a given chord instantly to just ratios. At the same time the global pitch is adjusted in such a way that the difference to the usual ET is minimized. This reduces the disturbing frequency shifts between subsequent chords. Hermode tuning also tries to compensate the so-called pitch drift (see section \ref{PitchDrift}).

\item\textit{TransFormSynth} is a freely available software-based synthesizer developed by William A. Sethares~\cite{Sethares}. Unlike all other approaches -- including the one presented here -- which are based on the idea of dynamically modifying the fundamental frequencies and thereby the whole series of corresponding partials, Sethares  proposes to keep the fundamental frequencies fixed (e.g. according to the ET). Instead, his algorithm modifies the frequencies of the higher partials in such a way that they match. As a result, even though the overtone spectra are distorted, the synthesized sound nevertheless tends to be perceived as consonant\footnote{A similar phenomenon occurs in the context of piano tuning. Since the overtone spectra of stiff steel strings are slightly inharmonic, piano tuners stretch the tuning in order to compensate these deviations.}. As far as we can see, this method is restricted to synthesizers which allow the overtone spectra to be specified individually, but it cannot be applied to ordinary musical instruments with natural harmonic overtone spectra.
\end{itemize}

\noindent
All these methods except for the last one are similar in that they analyze a given chord and then make decisions for tuning the frequencies, i.e. they are defined in procedural terms. Depending on the harmonic context, these decisions can be quite complex with different possible solutions for the same situation.

In the present paper, we investigate an alternative adaptive tuning scheme based on a different method which was originally proposed by~\cite{Laubenfels} (unpublished, for a short summary see~\cite{Sethares2005}). Instead of making a sequence of \textit{if-then} decisions, this method is defined mathematically and amounts to continuously solving a system of linear equations. As is described below, the system of equations may be viewed as a resistor network or likewise as a mechanical network of springs representing the interval sizes. Roughly speaking, each spring prefers to relax into a state where its length corresponds to the natural size of a pure interval and it will do so whenever this is possible, producing a chord in just intonation. However, if the spring network is so complex that it is impossible for all springs to simultaneously be situated in their ground state, the system will approach a non-trivial state under tension, representing a tempered harmonic compromise. This happens automatically without making any explicit decisions and may resemble the way in which musicians find the best possible intonation. As another advantage, the present method finds a harmonic compromise for \textit{all} intervals in a given chord, not only of the intervals between adjacent tones. 

To demonstrate the technical feasibility of the dynamically adjusting tuning scheme discussed in this paper, we implemented the tuning algorithm in an open-source application which is available for various platforms including mobile devices. This software will be described in more detail at the end of the paper.

\section{Definitions and Notations}

We start with basic definitions and notations, which will be used throughout this article.

\subsection{Frequencies and Equal Temperament}
In the following we consider a standard chromatic keyboard with keys enumerated from left to right by the index $k \in \{0,\ldots,K-1\}$.\footnote{For example, in the MIDI norm~\cite{MidiNorm} $k$ runs from 0 to 127 with the reference key A4 located at $k_0=69$.} For traditional keyboard instruments the corresponding frequencies $f_0,\ldots,f_{N-1}$ are constant throughout the performance, meaning that they have to be tuned beforehand according to a certain temperament. As mentioned above, Western music of today is predominantly based on the \textit{equal temperament} (ET) with twelve equally-sized semitones, defined by 
\begin{equation}
\label{fET}
f_k^\ET \;:=\; f_{k_0} \, 2^{(k-k_0)/12}\,,
\end{equation}
where $k_0$ denotes the index of the reference key and $f_{k_0}$ the corresponding reference frequency (usually A4 with $f_{k_0}=440$ Hz). 

An interval between two tones $k$ and $k'$ is characterized by a certain frequency ratio $f_{k'}/f_{k}$. In the ET this ratio is given by $f_{k'}^\ET/f_{k}^\ET = 2^{(k'-k)/12}$. The main advantage of the ET is that this ratio depends only on the difference $k'-k$, meaning that the frequency ratios are invariant under transpositions (key changes $k\to k+const$). Therefore, apart from the global pitch, the ET sounds identical in all keys. 

\subsection{Consonance and Just Intonation}
Two tones are justly intoned if the corresponding frequency ratio $f_{k'}/f_{k}$ is given by a simple rational number $R=m/n$. For example, a just octave has the frequency ratio 2:1 while the just perfect fifth corresponds to the ratio 3:2 (see Table \ref{tablejustint}). As outlined in the introduction, the impression of consonance is particularly pronounced if $m$ and $n$ are small. 

\textit{Just intonation} (JI) is a tuning scheme where the frequencies $f_k$ are tuned according to rational numbers with respect to a certain keynote $k^*$ in an octave-repeating pattern:
\begin{equation}
f^\JI_k = R^\JI_{k,k^*} \, f^\JI_{k^*}\,.
\end{equation}
A possible choice of the rational numbers $R^\JI_{k,k^*}$ is given in Table \ref{tablejustint}. With respect to the keynote the resulting interval ratios $f^\JI_k/f^\JI_{k^*}$ are exactly those listed in the table. However, in contrast to the ET these frequency ratios are not invariant under transpositions. For example, for keynote C the fifth C-G has the correct frequency ratio 3:2 but the fifth D-A has the ratio $40/27 \simeq 1.48$ which is clearly too small. Therefore, as already outlined in the introduction, JI as a static tuning can only be used in the scale referring to its keynote (and a few complementary keys) while it sounds dissonant in most other keys. 

\begin{table}
\centering
\footnotesize
\begin{tabular}{|c|c||c|c||c|c|c|} \hline
$k'-k$ & Interval& $f^\ET_{k'}/f^\ET_{k}$ & $R^\JI_{k,k'}=m/n$ & $\Phi^\JI_{k,k'}[\cent]$ & $\Phi^\ET_{k,k'}[\cent]$ & Deviation  $\phi^\JI_{k,k'}[\cent]$ \\ \hline 
0 & Unison 		& 1		& 1 	& 0 & 0 & 0 \\
1 & Semitone 		& 1.0595	& 16/15 & 111.73 & 100 & +11.73 \\
2 & Major Second 	& 1.1225 	& 9/8 	& 203.91 & 200 & +3.91 \\
3 & Minor Third 	& 1.1892	& 6/5 	& 315.64 & 300 & +15.64 \\
4 & Major Third 	& 1.2599	& 5/4 	& 386.31 & 400 & -13.69 \\
5 & Fourth 		& 1.3348 	& 4/3 	& 498.04 & 500 & -1.96 \\
6 & Tritone 		& 1.4142	& 45/32 & 590.22 & 600 & -9.78 \\
7 & Fifth 		& 1.4983	& 3/2 	& 701.96 & 700 & +1.96 \\
8 & Minor Sixth 	& 1.5874	& 8/5 	& 813.69 & 800 & +13.69 \\
9 & Major Sixth 	& 1.6818	& 5/3 	& 884.36 & 900 & -15.64 \\
10 & Minor Seventh 	& 1.7818	& 9/5 	& 1017.60 & 1000 & +17.60 \\
11 & Major Seventh 	& 1.8878	& 15/8	& 1088.27 & 1100 & -11.73 \\
12 & Octave 		& 2		& 2 	& 1200 & 1200 & 0 \\
\hline 
\end{tabular}
\caption{\label{tablejustint} List of chromatic interval sizes. The table shows the number of semitones $k-k'$, the frequency ratio $f^\ET_{k'}/f^\ET_{k}$ in the equal temperament, the just ratios $R^\JI_{k,k'}=m/n$ according to Table~\ref{tab:frequencyratios}, the interval sizes $\Phi_{k,k'}$ in cents for just intonation (JI) and the equal temperament (ET), as well as the cent difference $\phi^\JI_{k,k'}$ between the two values. Note that the choice of $m/n$ for a given interval is not always unique. For example, the minor seventh can be tuned according to the ratio 9/5 or 7/4 (see Table \ref{tableseveralintervalstablejustint} in Sect. 4).}
\end{table}

\subsection{Logarithmic pitches and interval sizes}
Combining several musical intervals in a chord the corresponding frequency ratios multiply. Therefore, as already pointed out by Huygens in 1691 \cite{Huygens,Cohen}, it is convenient to consider the \textit{logarithm} of frequency ratios, quantified in units of cents. The logarithm transforms multiplication into addition and allows one to add the sizes of adjacent intervals. Using this convention we define pitches and interval sizes as follows:
\begin{itemize}
\item 
  The \textit{absolute pitch} $\Lambda_k$ of the key $k$ on the keyboard is defined as the cent difference between the frequency of the key $k$ and that of the reference key $k_0$ (A4):
  \begin{equation}
  \Lambda_k \;:=\; 1200 \log_2 \Bigl( \frac{f_k}{f_{k_0}}\Bigr)=1200\Bigl(\log_2 f_k- \log_2 f_{k_0}\Bigr) \,,
  \end{equation}
  where $\log_2 f =\log f/\log 2$ denotes the logarithm to the base 2. For example, the pitches in the ET~(\ref{fET}) are simply given by multiples of 100 cent:
  \begin{equation}
  \Lambda^\ET_k \;=\; 100(k-k_0).
  \end{equation}
  For the actual \textit{pitch deviation} of the key $k$ relative to the ET we use the notation
  \begin{equation}
  \lambda_k \;:=\; \Lambda_k - \Lambda^\ET_k \,.
  \end{equation}
  \vspace{4mm}

\item 
  The microtonal \textit{absolute interval size} $\Phi_{k,k'}$ between two keys $k$ and $k'$ is defined as the corresponding pitch difference in units of cents:  
  \begin{equation}
  \Phi_{k,k'} \;=\; \Lambda_{k'}-\Lambda_k \;=\; 1200\Bigl(\log_2 f_{k'}- \log_2 f_k\Bigr).
  \end{equation}
  In the ET~(\ref{fET}) the interval sizes are given by the number of semitones times 100:
  \begin{equation}
  \Phi^\ET_{k,k'} \;=\; \Lambda^\ET_{k'}-\Lambda^\ET_k \;=\; 100(k'-k).
  \end{equation}
  For the actual \textit{interval size deviation} from the ET we use the notation
  \begin{equation}
  \phi_{k,k'} \;:=\; \Phi_{k,k'} -\Phi^\ET_{k,k'} \;=\; \lambda_{k'}-\lambda_k \,. 
  \end{equation}
  For JI a list of possible values for $\Phi^\JI_{k,k'}$ and $\phi^\JI_{k,k'}$ is given in Table \ref{tablejustint}.
\end{itemize}
%

\section{Vertical Intonation -- Adaptive Tuning of a Single Chord}

Adaptive tuning schemes are confronted with two important aspects of tuning. On the one hand, each new chord has to be intoned `vertically', that is, one has to tune the relative pitches between simultaneously played notes. On the other hand, subsequent chords have to be intoned relative to each other in `horizontal' (temporal) direction according to the harmonic progression, as will be discussed in the following section.

\subsection{Vertical tuning at first glance}

In order to tune a given chord vertically, we want to determine the pitches in such a way that the resulting interval sizes are equal or at least close to the ideal ratios of JI. In other words, for a chord consisting of $N$ tones according to the keyboard keys $\{k_1,k_2,\ldots,k_N\}$ we have to find pitches $\{\Lambda_{k_1},\ldots,\Lambda_{k_N}\}$ such that the interval sizes $\Phi_{k_i,k_j}$ agree as much as possible with the values~$\Phi^\JI_{k_i,k_j}$ listed in Table \ref{tablejustint}.

Most of the existing approaches mentioned in the introduction consider only the intervals between \textit{adjacent} tones of a chord. Contrarily the present method also takes intervals between non-adjacent tones into account on equal footing with the others. This means that for a chord consisting of $N$ tones there are $N(N-1)/2$ possible intervals which have to be tuned as just as possible.

As there are $N(N-1)/2$ intervals but only $N$ degrees of freedom, it is clear that it is not always possible to find a consistent solution where all interval sizes $\Phi_{k_i,k_j}$ match exactly with the given cent differences $\Phi^\JI_{k_i,k_j}$. For example, the triad C-E-G can be tuned in just intonation while the triad C-E-G$^\sharp$ cannot.\footnote{C-E-G: Tuning the major third C-E to the ratio 5/4 and the minor third E-G to the ratio 6/5, the resulting fifth C-G has the  ratio $\frac54\cdot \frac65 = \frac32$, meaning that the triad is perfectly consonant in just intonation. Contrarily, for C-E-G$^\sharp$, where two major thirds are combined, the resulting augmented fifth would have the frequency ratio $\bigl( \frac54\bigr)^2=\frac{20}{16}$ which differs from the ratio $\frac85$ listed in the Table \ref{tablejustint}. This means that this chord cannot be tuned consistently in just intonation.} In such a situation, where a chord cannot be tuned consistently in just intonation, the algorithm should render an acceptable tempered compromise. In fact, this is basically what musicians do: they do not solve complicated puzzles of number theory, instead they simply adjust their own pitch on an intuitive basis in such a way that the best possible harmonic compromise is achieved.

\begin{figure}[t]
\fl\begin{minipage}{110mm}
\caption{\label{circuit}Simplified sketch of the vertical tuning scheme suggested in the present paper. The figure shows a keyboard on which a C-Major chord is played. Viewing these keys as electrical contacts we place a battery in series with a resistor between each of the $4\cdot 3/2=6$ possible intervals. Assuming that each battery has a voltage equal to the ideal pitch difference $\Phi^\JI_{k_i,k_j}$ in JI, the resistor network will attain an equilibrium according to Kirchhoff's laws where the voltages at the key contacts represent the desired microtonal pitches. If all electrical currents in the network vanish (as in the present example) the chord is tuned exactly in JI. If not, the specific choice of the resistors determines a tempered compromise where the dissipated power measures how much the chord is tempered. The system is coupled to an external voltage  which controls the reference pitch.} 
\end{minipage}
\vspace{-76mm}
\begin{flushright}
\includegraphics[width=60mm]{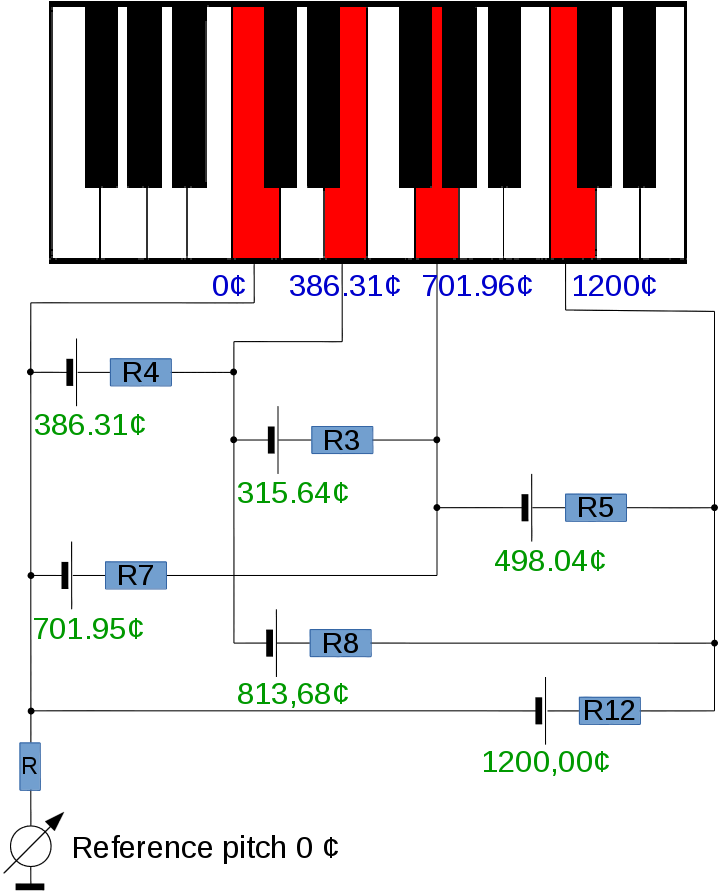} 
\end{flushright}
\end{figure}

The solution investigated in this paper is based on a simple idea which can be explained as follows. As sketched in Fig.~\ref{circuit}, we consider a fictitious battery-resistor network, where each battery has a voltage corresponding to the ideal pitch difference $\Phi^\JI_{k_i,k_j}$ of JI. If the chord can be tuned justly (e.g. as a major triad) the voltages will adjust exactly at the corresponding pitches and the currents passing the resistors are zero. Otherwise, for chords which cannot be tuned justly, the network will produce a certain compromise which depends on the specific choice of the resistors. As already pointed out above, the dissipated power can be regarded as a measure how strongly this tuning compromise is tempered.

\subsection{Mathematical formulation}
Consider a chord of $N$ tones with key indices $k_1,k_2,\ldots,k_N$ in increasing order. The chord consists of $N(N-1)/2$ intervals with index pairs $i,j \in \{k_1,\ldots,k_N\}$ ordered by $i<j$. The task would be to tune the pitches $\Lambda_i$ (with $i\in\{k_1,\ldots,k_N\}$) in such a way that the pitch differences $\Lambda_j-\Lambda_i$  deviate as little as possible from the ideal pitch differences $\Phi^\JI_{k_i,k_j}$ listed in Table \ref{tablejustint}, or equivalently, that the differences $\lambda_j-\lambda_i$ deviate as little as possible from $\phi^\JI_{i,j}:=\phi^\JI_{k_i,k_j}$. We solve this problem by minimizing the squared deviations as follows. Denoting by $\vec\lambda=(\lambda_{k_1},\ldots,\lambda_{k_N})^T$ the vector of pitch deviations from the ET of the pressed keys, we define a deviation potential by
\begin{equation}
\label{V1}
V[\vec\lambda] \;=\; \frac14 \sum_{i,j\in \{k_1,\ldots,k_N\} \atop i<j} w_{ij} \Bigl( \lambda_j(t)-\lambda_i(t)-\phi^\JI_{ij}\Bigr)^2 \,,
\end{equation}
which is just the sum of all quadratic deviations of the interval sizes weighted by certain factors $w_{ij}$, assuming that $w_{ii}=0$. The weights can be chosen freely and can be viewed as the conductivity of the resistors in Fig.~\ref{circuit}. Their purpose is to determine the `rigidity' of the respective interval in the tuning process. In practice it is meaningful to assign a high weight factor to intervals with simple fractional ratios. In addition, the weight factors may also take the actual volume of the participating tones into account. 

In a more compact notation, the deviation potential (\ref{V1}) can be written in the bilinear form
\begin{equation}
\label{GeneralV}
V[\vec\lambda] \;=\; \frac12 \vec\lambda\cdot \mathbf{A} \vec\lambda \;+\; \vec b\cdot \vec\lambda + c\,,
\end{equation}
where $\mathbf{A}$ is a symmetric $N\times N$ matrix and $\vec b$ is a vector with the components
\begin{equation}
\label{AB}
A_{ij} \;=\; \left\{
\begin{array}{cc}
-w_{ij} & \mbox{ if } i\neq j  \\
\sum_{\ell} w_{i\ell} & \mbox{ if } i=j 
\end{array}
\right.\,,\qquad
b_i \;=\; \sum_j w_{ij} \,\phi_{ij}
\end{equation}
and where
\begin{equation}
\label{C}
c=\frac14 \sum_{i,j} w_{ij} \phi_{ij}^2
\end{equation}
is a constant. The optimal pitches $\vec\lambda^\OPT$, in which we would like to tune the chord, correspond to a situation where $V[\vec\lambda]$ is minimal, that is $\vec\nabla_\lambda V=\vec 0$, leading to the system of equations $\mathbf{A}\vec\lambda +\vec b=0$. Thus, if $\mathbf{A}$ was invertible, the solution would be given by
\begin{equation}
\vec\lambda^\OPT=-\mathbf{A}^{-1}\vec b.
\end{equation}
Thus, the whole tuning process amounts to solving a system of linear equations. Finally, the potential 
\begin{equation}
\label{eqv}
V[\vec\lambda^\OPT] \;=\; c-\frac12 \vec b \cdot \mathbf{A}^{-1}\vec b
\end{equation}
evaluated at $\vec\lambda=\vec\lambda^\OPT$ gives the dissipated power, telling us to what extent the result is tempered.

However, inspecting $\mathbf A$ one can easily see that the column sum is zero, hence the matrix does not have full rank and thus it is not invertible. This can be traced back to the fact that the potential is defined in pitch \textit{differences}, leaving the absolute pitch of the chord undetermined. This can be circumvented easily by coupling the network to an external source which determines the global concert pitch, as described in Appendix~A.

\section{Horizontal Intonation -- Adaptive Tuning in the Harmonic Progression}

Normally we perceive combinations of tones as `in tune' or `out of tune' if they are played simultaneously. However, as demonstrated by~\cite{milne2016}, we are also capable of recognizing the intonation of sequentially played tones, provided that the elapsed time  in between is not too large. Apparently our sense of hearing is able to memorize sounds and their spectra for a short while. In empirical studies it was found that this psychoacoustic intonational short-term memory is characterized by a typical time scale of about three seconds~\cite{WittmannPoeppel,LehmannGoldhahn}. 

If the chords are tuned separately as described in the last section, the sudden change of the chordal root may lead to unpleasant intonational discontinuities between chords. This requires that the intonational memory has to be taken into account by correlating the pitches of subsequent chords in a harmonic progression. In the following we describe how the intonational memory can be incorporated in the suggested framework of adaptive tuning.

\subsection{Intonational memory}

Pressing a key $k$ the instrument produces a sound with the time-dependent intensity (volume) $I_k(t)$ which decays to zero when the key is released. In order to implement the intonational memory, we introduce a \textit{memory function} $M_k(t)$ interpreted as the `virtual intensity' at which the sound of a key $k$ is memorized. When a new key is pressed $M_k(t)$ is initially set to the actual intensity $I_k(t)$. Thereafter, it follows $I(t)$ by means of the over-damped first-order differential equation
\begin{equation}
\frac{\d M_k(t)}{\d t} \;=\; \frac{1}{\tau_M} \, \Bigl(I_k(t)-M_k(t)\Bigr)\,,
\end{equation}
where $\tau_M \approx 3$s is the characteristic time scale of the intonational memory. For example, if the volume of a key drops suddenly to zero after releasing a key, $M(t)$ will decrease exponentially as $e^{-t/\tau_m}$.

This simple model of the intonational memory can be improved further by observing that it also takes some time to recognize the pitch of a newly pressed key. In fact, it is quite easy to memorize the pitches of long sustained notes while individual short notes at high tempo are much harder to remember. This suggests that there is another typical time scale $\tau_R$ for recognizing the pitch of a sound which may be taken into account by considering the dynamics
\begin{equation}
\label{memory}
\frac{\d M_k(t)}{\d t} \;=\; 
\left\{ \begin{array}{cc}
        \frac{1}{\tau_R} \Bigl(I_k(t)-M_k(t)\Bigr) & \mbox{ if } M_k(t) < I_k(t) \\[2mm]
        \frac{1}{\tau_M} \Bigl(I_k(t)-M_k(t)\Bigr) & \mbox{ if }  M_k(t) \geq I_k(t)
        \end{array}\,.
\right.
\end{equation}
Among musicians the typical value of $\tau_R$ is expected to be smaller than $\tau_M$, and it seems that values in the vicinity of one second are a reasonable choice.

\subsection{Horizontal adaptive tuning}

In order to correlate the intonation of subsequent chords, we use the same mechanism as described above for the case of vertical tuning. To this end let us consider a memorized key with the index $k_m$ that was tuned to the pitch $\tilde \Lambda_m$, followed by a newly pressed key with the index $k_i$ (including the case that the same key is pressed again). The aim is to tune the pitch $\Lambda_i(t)$ dynamically in such a way that the interval size $\tilde\Lambda_m-\Lambda_i(t)$ approximates as much as possible the ideal interval size $\Phi_{k_i,k_m}^\JI$ of JI, as listed in Table \ref{tablejustint}. In other words, we have to determine $\vec\lambda$ in such a way that $\tilde\lambda_m-\lambda_i(t)$ deviates as little as possible from $\phi_{im}^\JI:=\phi_{k_i,k_m}^\JI$. This leads to simply adding a memory term in the potential

\begin{equation}
 \fl
V[\vec\lambda] \,=\; \frac14 \sum_{i,j} w_{ij}(t) \Bigl( \lambda_j(t)-\lambda_i(t)-\phi^\JI_{ij}\Bigr)^2 \,+\,
\frac{1}{2} \sum_{i}\sum_{m}\tilde w_{im}(t)\, \Bigl( \tilde\lambda_m-\lambda_i(t)-\phi^\JI_{im}\Bigr)^2
\end{equation}
where $k_i,k_j$ with $i,j\in \{1,\ldots, N\}$ run over all audible keys while $k_m$ with  $m\in\{1,\ldots,M\}$ runs over the memorized keys. Here $\tilde w_{im}(t)$ is a time-dependent weight factor reflecting the actual intensity of the key $k_i$ and the memorized intensity of the key $k_m$. Again this potential can be written in the vector notation (\ref{GeneralV}) with
\begin{eqnarray}
\label{eqa}
\mathbf{A}_{ij} &=& \left\{
\begin{array}{ll}
-w_{ij} & \mbox{ if } i\neq j  \\
\sum_{\ell} w_{i\ell} \,+\, \sum_m \tilde w_{im} & \mbox{ if } i=j 
\end{array}\right.
\\[2mm]
\label{eqb}
b_i &=& \sum_j w_{ij} \,\phi_{ij} \;+\; \sum_m \tilde w_{im} \,
(\phi_{im}-\tilde\lambda_m) \\
\label{eqc}
c&=&\frac14 \sum_{i,j} w_{ij} \phi_{ij}^2 \,+\, \frac{1}{2} \sum_i \sum_m \tilde w_{im} \,
(\phi_{im}-\tilde\lambda_m)^2
\end{eqnarray}
and its minimum is attained at $\vec\lambda^\OPT=-\mathbf A^{-1}\vec b$. Note that this method automatically finds a tempered compromise if the chordal roots of a harmonic progression are incompatible.

\subsection{Pitch Drift Compensation}
\label{PitchDrift}

One of the major disadvantages of dynamical tuning schemes with temporal correlation is the gradual migration of the overall pitch. For example, playing a full chromatic scale of 12 semitones with  fixed sizes $\Phi_{k,k+1}^\JI$=111.73$\cent$ (frequency ratio $16/15$) one ends up with 1340.76$\cent$, which is more than a half tone higher than an octave. In practice the pitch drift meanders in positive and negative direction, depending on the actual harmonic progression.

The pitch drift can be reduced by admitting different interval sizes, as will be explained in the next section. For example, 12  semitones with alternating sizes of $111.73\cent$ $(16/15)$ and $92.18\cent$ $(135/128)$ form six whole tones of $3.91\cent$ ($9/8$) add up to $1223.46\cent$ which is much closer to a just octave of $1200\cent$. But even this improvement does yet not eliminate the pitch progression entirely.

It is therefore meaningful to implement an additional pitch drift compensating mechanism by slowly adjusting all pitches uniformly in such a way that the desired reference pitch is approached, as described by the differential equation
\begin{equation}
\frac{\d \lambda_j}{\d t} \;=\; \frac{1}{\tau_c} \,
\Biggl(\Bigl(\frac{1}{N}\sum_{i\in\{k_1,\ldots,k_N\}} \lambda_i\Bigr)-\lambda^\REF \Biggr)\,,\qquad {j \in\{k_1,\ldots,k_N\}\,.
}\end{equation}
Here $\lambda^\REF=1200*\log_2(f_{k_0}/440$ Hz) is the global pitch versus 440 Hz while $\tau_c$ defines the typical time scale on which the compensation takes place.  Note that the pitch drift affects the frequencies of all pressed keys equally without changing the relative frequency ratios between them. This means that the harmonic texture of the sound remains the same, only the overall pitch varies slowly with time. In contrast to the vertical and horizontal tuning, which takes place immediately after pressing a new key, the time scale for the pitch compensation is much larger and should be chosen such that the gradual compensation not noticeable for the listener.

\section{Dealing with Non-Unique Interval Sizes}

So far the method outlined above determines the best possible tuning result for a fixed table of given interval sizes (see Table \ref{tablejustint}). However, the frequency ratios in JI are not unique; rather there are various possible choices for certain intervals, defining different variants of just intonation (see Table \ref{tableseveralintervalstablejustint}). This suggests that the tuning result can be improved by finding the best possible solution among these variants~\cite{Duffin}.

\begin{table}
\centering
\footnotesize
\begin{tabular}{|c|c||l|l|l|} \hline
$k'-k$ & Interval Name & \multicolumn{3}{|c|}{Tuning alternatives $p/q$ \ \ \ ($\phi^\JI_{k,k'}[\cent]$)} \\ \hline
0 & Unison 		& 1/1 (0)	&		&		\\
1 & Semitone 		& 16/15 (+11.73)& 25/24 (-29.32)&		\\
2 & Major Second 	& 9/8 (+3.91)	& 10/9 (-17.60)	&		\\
3 & Minor Third 	& 6/5 (+15.64)	&		&		\\
4 & Major Third 	& 5/4 (-13.69)	&		&		\\
5 & Fourth 		& 4/3 (-1.96)	&		& 		\\
6 & Tritone 		& 45/32 (-9.78) &		&		\\
7 & Fifth 		& 3/2 (+1.96)	&		&		\\
8 & Minor Sixth 	& 8/5 (+13.69)	&		&		\\
9 & Major Sixth 	& 5/3 (-15.64)	&	 	& 		\\
10 & Minor Seventh 	& 16/9 (-3.91)	& 9/5 (+17.60)	& 7/4 (-31.17)	\\
11 & Major Seventh 	& 15/8 (-11.73) &		&		\\
12 & Octave 		& 2/1 (0)	&		&		\\
\hline 
\end{tabular}
\caption{\label{tableseveralintervalstablejustint} List of possible frequency ratios of just chromatic intervals. The third column contains the same data as in Table ~\ref{tablejustint}. In addition the most important alternative tuning ratios are shown.}
\end{table}

The advantage of admitting several variants can be explained by the following example. Two successive major seconds with the ratio of 9/8 (+3.9\cent) make up a major third with the ratio of 81/64 ($+7.8\cent$) which differs significantly from the just ratio of 5/4 ($-13.7\cent$). However, if we combine two \textit{different} justly intoned variants of major seconds with the ratios of 9/8 ($+3.9\cent$) and 10/9 ($-17.6 \cent$) the resulting major third has exactly the just frequency ratio of 5/4 ($-13.7\cent$).

What determines the correct choice of the interval size? In a procedural setting this is a highly complex music-theoretical problem that requires a thorough analysis of the harmonic progression. But in practice the decision for the best fitting interval is made aurally on an intuitive basis (although a general understanding of the harmonic progression is part of the process). Inspired by this observation we implement non-unique interval sizes in the algorithm described above by repeating the minimization procedure for all possible combinations of the alternative interval sizes listed in Table~\ref{tableseveralintervalstablejustint} and then to take the solution with the lowest deviation from JI. 

The four alternative ratios listed in Table~\ref{tableseveralintervalstablejustint} have been chosen empirically. There are of course many more possible ratios which could be added as well. However, as we minimize over all possible combination of interval sizes, it is clear that the execution time grows exponentially with the number of alternative ratios. This is the reason why the table is restricted to only four alternative entries for the most flexible intervals.

\section{Open-Source Demonstration Software -- Technical Details}

\begin{figure}[b]
\centering\includegraphics[width=100mm]{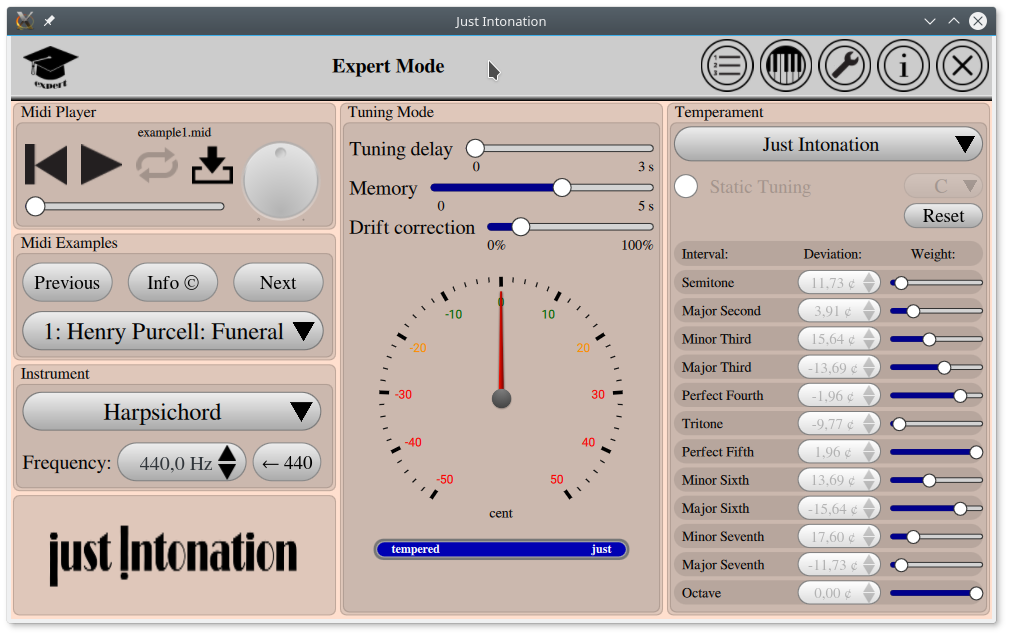}
\caption{\label{JIapp}
Snapshot of the application \textit{Just Intonation} in the expert mode.
}
\end{figure}

In order to demonstrate the tuning method discussed in this paper, we initiated an open-source project called~\textit{\cite{JustIntonation}} (the present paper refers to version 1.3.2).  This software allows the user to hear and play music with and without adaptive tuning.  Audio examples, a short video and download links for various platforms are available on our website \href{http://www.just-intonation.org}{\texttt{www.just-intonation.org}}.

The application  has been designed as an educational application rather than a professional tool for producing music. It provides a \textit{simple mode} for getting started as well as an \textit{expert mode} for more sophisticated experiments (see screenshot in Fig.~\ref{JIapp}).

\begin{figure}
\centering\includegraphics[width=130mm]{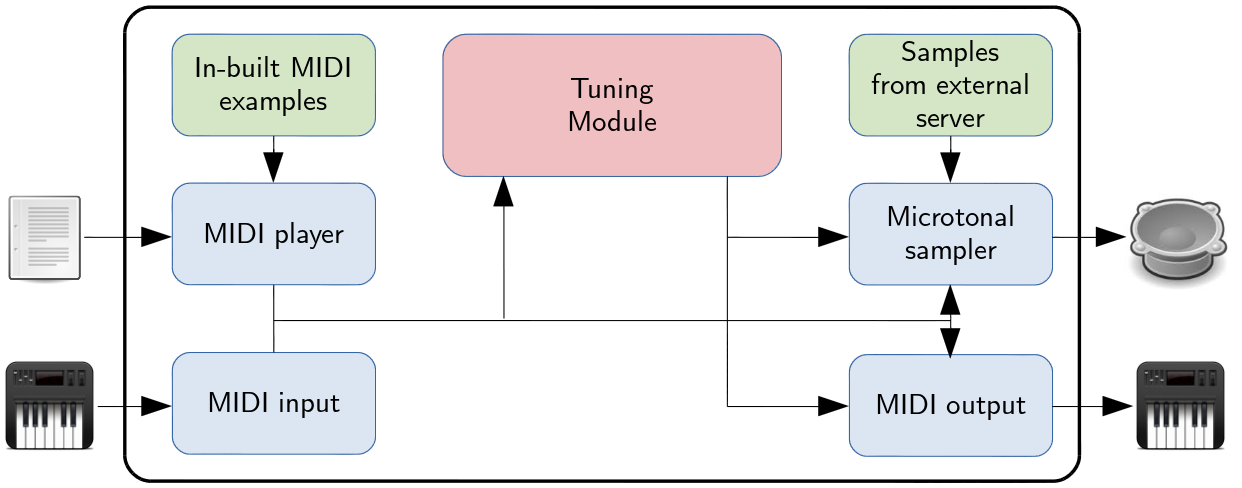}
\caption{\label{appstructure}
Basic structure of the application and its modules.}
\end{figure}

\textit{Just Intonation} is a multi-platform application for desktop computers and mobile devices written in C++  based on Qt\texttrademark\footnote{Qt\texttrademark ~is a cross-platform application framework licensed under LGPL that is used for developing application software, see \href{https://www.qt.io/}{\tt www.qt.io/}}. As sketched in Fig.~\ref{appstructure}, it contains various submodules which partially run in different threads and communicate with each other via Qt signals. MIDI messages generated by an integrated player or an external device are sent to the tuning module and to the sound-generated modules. Depending on the MIDI data the tuning module continually computes the vector $\vec\lambda^\OPT$ according to the formulas given above and and emits the calculated pitches via Qt signals to the audio modules. The application includes an built-in microtonal sampler which can play triangular waves as well as realistic samples recorded by the authors (piano, organ, and harpsichord). As the application is designed primarily for educational purposes, there is no particular emphasis on low-latency audio. 

Alternatively, it is possible to connect an external MIDI device which is capable of processing 'pitch-bends'. Normally the MIDI pitch-bend command modifies the frequencies of all depressed keys uniformly. In order to circumvent this restriction, the MIDI output module remaps the incoming MIDI stream to 15 different channels, tuning each of them individually via pitch-bend. Note that this restricts the output to a 15-voice polyphony of a single instrument. It should be mentioned at this point that the MIDI standard has recently been extended. This new standard, called ``MIDI Polyphonic Expression''~\cite{MPE}, overcomes the aforementioned limitations and allows for a direct microtonal control of the individual pitches. 

\begin{figure}
\centering\includegraphics[width=150mm]{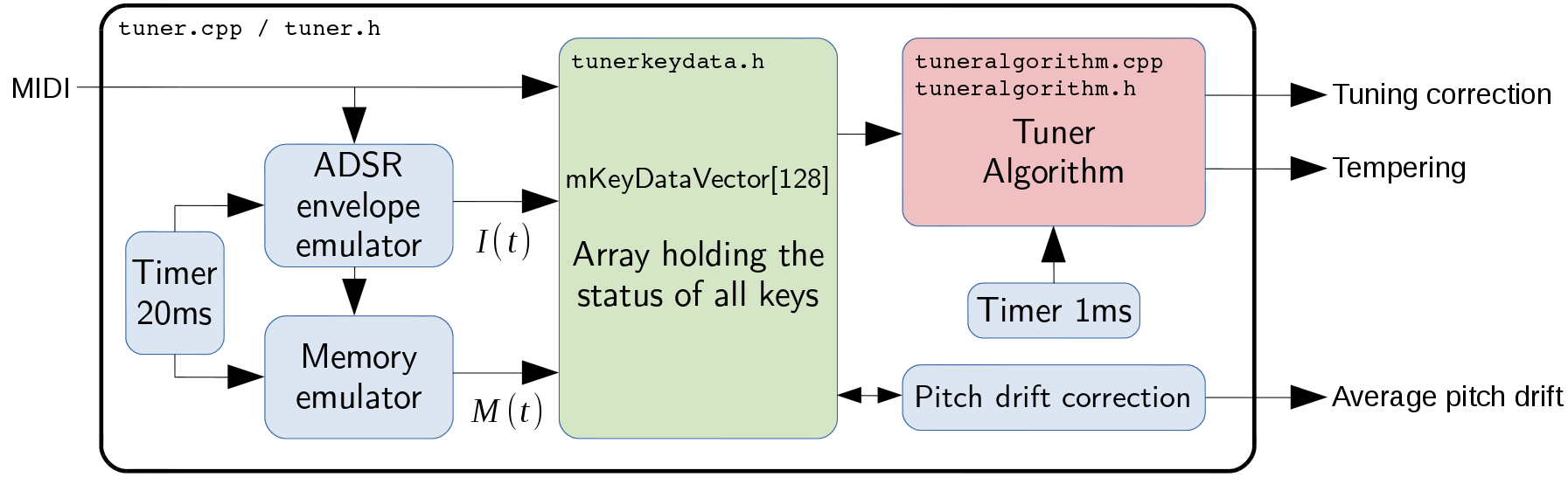}
\caption{\label{tunermodule}
Internal structure of the tuner module.}
\end{figure}

The main module of interest is the tuning module. This module runs entirely in a separate event loop of an independent thread and communicates with the application via Qt signals, ensuring thread safety. Its internal structure is shown in Fig.~\ref{tunermodule}. Its main functionality is 
\begin{itemize}
\item receiving MIDI signals and sending tuning signals,
\item emulating the intensity $I(t)$ as well as the memory $M(t)$ for each key of the keyboard,
\item keeping track of the status of the keys (including on/off, volume and basic envelope) in an array of type $\tt KeyData$,
\item executing the {\tt TunerAlgorithm} every 20ms or upon incoming MIDI events, and 
\item managing the pitch drift compensation.
\end{itemize}

For further technical details and code excerpts the interested reader is referred to Appendix~B.

\section{Outlook}

The development of temperaments and the ongoing tug-of-war between just tuning and transposability over many centuries is a fascinating aspect in the history of music theory and practice. At the beginning of the 20$^{\rm th}$ century it seemed as if the universal acceptance of the ET had finally settled this issue and, in fact, still today this is the prevailing view. In contrast, we share the opinion that the quest for a better intonation is not over yet and that the ET is probably only an intermediate rather than a final solution.

Looking back at the past 150 years it seems that the search for a better intonation oscillates between enthusiasm and disillusionment. For example, starting with the seminal work by Helmholtz~\cite{helmholtz} in 1865, who was among the first to provide a scientific basis for the sensation of music, many theorists and instrument makers at the end of the 19$^{\rm th}$ century were inspired by the challenge to construct a ``\textit{Reininstrument}'', but the solutions were simply too complicated to become accepted on a broad scale. Then in the early 20$^{\rm th}$ century the interest in just intonation abated, probably in part because many composers were writing music that was highly dissonant and even atonal.

In the second half of the last century, a  renewed interest in just intonation and different ways of tuning arose alongside with an increasing attention on historical performance practices~\cite{DuffinArticle}. The new technologies becoming available constituted  another promoting factor for this process. Following the visionary contributions by Eivind Grove~\cite{GrovenMax}, the emerging computer technology led to a variety of proposals, patents, and software packages which reflect the technological capabilities of the respective time. Unfortunately, apart from few exceptions, none of these approaches reached a broader dissemination, partly because the whole issue had maintained the reputation of being exotic and academic, linked to the microtonal community where 12 tones per octave are considered merely as an exception rather than a rule. 

Meanwhile, however, an ordinary mobile phone has become more powerful than a supercomputer in the 80's, offering new and previously undreamt-of possibilities. For example, solving a system of equations in real time, as proposed in the present work, would have been inconceivable two decades ago. Moreover, digital information technology continues to change the musical landscape and the art of instrument making entirely. The purpose of this project is to demonstrate that by now we have completely different means at our disposal which allow us to consider different approaches and to make dynamic tuning schemes suitable for everyday use.  A systematic evaluation of the tuning method is beyond the scope of this article, but we invite the interested readers to form their own subjective conclusions by testing the freely available software or by listening to the available sound examples on our website \href{http://www.just-intonation.org}{\texttt{just-intonation.org}}.

Finally, electronic communication increasingly enhances the interaction between different musical cultures. On the one hand, it is obvious that many traditional intonation systems throughout the world are increasingly influenced (if not even destroyed) by the Western ET. On the other hand, it should not be underestimated that this interaction also influences the Western world, and it cannot be ruled out that at some point in the future it may become fashionable to deviate from the ET. In addition, it is to be expected that the art of instrument making will continue to evolve rapidly and that on the long run the importance of statically tuned temperaments may decrease. All this suggests that dynamically adapting tuning scheme might become more important in the future. This does not necessarily mean that just intervals are the ultimate goal, and in fact it has been shown by~\cite{mathews} that small deviations from rational frequency ratios may certainly perceived as pleasant, but perhaps there will be a growing interest in more consonant intonation schemes. With our contribution we would like to point out that there is a lot of open space for further research and development in this direction.

\vspace{2mm}
\noindent
\textbf{Acknowledgements}\\
We would like to thank the anonymous reviewers for their substantial reports and in particular for making us aware of the work by John de Laubenfels. We also thank R. W. Duffin for the stimulating exchange of ideas, in particular for his suggestion to consider variable interval sizes. We are also grateful to U. Konrad, who made it possible to record the samples for the application. Finally, we would like to thank A. Whisnant for critical reading of the manuscript.

\appendix
\section{Controlling the Global Pitch}

To solve the problem of non-invertibility we couple the network weakly to the global reference pitch, as indicated at the bottom of Fig.~\ref{circuit}. This amounts to adding the term to the potential~(\ref{GeneralV})
\begin{equation}
V[\vec\lambda] \;=\; \frac14 \sum_{i,j=1}^N w_{ij} \Bigl( \lambda_j(t)-\lambda_i(t)-\phi^\JI_{ij}\Bigr)^2 \,+\,
\frac \epsilon 2 \sum_{i=1}^N ( \lambda_i - \lambda_\REF)^2\,,
\end{equation}
where $\lambda_\REF = 1200 \log_2 (f_{k_0}/440$ Hz) is the deviation from the reference pitch and
$\epsilon \ll w_{ij}$ is a small coupling parameter, replacing Eqs. (\ref{AB}) and (\ref{C}) by the modified expressions
\begin{eqnarray}
\mathbf{A}_{ij} &=& \left\{
\begin{array}{cc}
-w_{ij} & \mbox{ if } i\neq j  \\
\epsilon + \sum_{\ell} w_{i\ell} & \mbox{ if } i=j 
\end{array} \right.
\\
b_i &=& - \epsilon \lambda_\REF + \sum_j w_{ij} \,\phi_{ij}
\\
c&=&\frac14 \Bigl( 2N \epsilon \lambda_\REF^2 + \sum_{i,j} w_{ij} \phi_{ij}^2\Bigr) \,.
\end{eqnarray}
Now the matrix $\mathbf A$ is invertible and the optimal pitch differences can be computed by $\vec\lambda^\OPT=-\mathbf{A}^{-1}\vec b$. A similar modification can be made if horizontal tuning is taken into account.

\section{Appendix: Tuning Module -- Technical Details}

In the source code, which can be downloaded from \href{https://gitlab.com/tp3/JustIntonation}{\tt https://gitlab.com/tp3/JustIntonation}, the core of the tuning module can be found in the directory {\tt ./application/modules/tuner}. The tuner interface is realized as an instance of the class {\tt Tuner} which is in turn running an instance of the {\tt TunerAlgorithm} in an independent thread. 

\begin{figure}[t]
\centering\includegraphics[width=150mm]{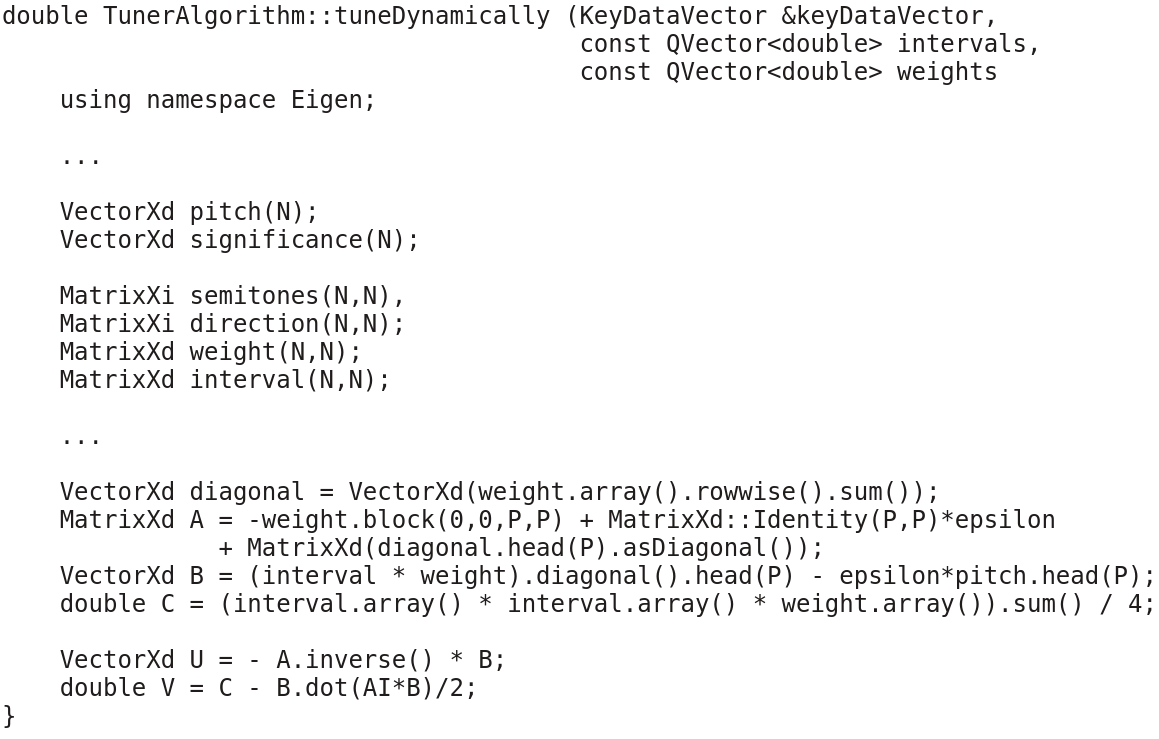}
\caption{Code excerpt from the tuning module (see text).\label{code}}
\end{figure}

A greatly shortened excerpt of the main algorithm without optimization is shown in Fig.~\ref{code}. As can be seen, the function {\tt tuneDynamically} is called with three arguments, passing a structure containing the status of all keys, a vector of the desired interval sizes in cents, and a vector of the corresponding weights which can be adjusted individually in the expert mode of the application.

The implementation of the tuner algorithm is based on the open-source library,\footnote{Open-source software library for linear algebra, see \href{http://eigen.tuxfamily.org}{\tt eigen.tuxfamily.org}} a C++ library for linear algebra, which is used here to solve the system of linear equations described above. First two vectors and four matrices are declared and initialized as follows (the initialization is not shown in Fig.~\ref{code}):
\begin{itemize}
 \item {\tt pitch} is a vector containing the actual pitches of the pressed keys in cents.
 \item {\tt significance} is a vector holding the weight of each pressed key depending on its volume.
 \item {\tt semitones} is a matrix counting the number of semitones between each pair of the pressed keys.
 \item {\tt direction} indicates whether the corresponding interval goes up or down.
 \item {\tt weight} is an array containing the tuning weights according to the slider setting in the expert mode.
 \item {\tt interval} contains the desired JI interval sizes in cents.
\end{itemize}
After initializing these objects, we use the \textit{Eigen} library to set up $A,B$, and $C$ (see Fig.~\ref{code}) and finally we solve Eqs. (\ref{eqa})-(\ref{eqc}). The actual solution of the problem is carried out in a single line, namely

\begin{center}
{ \tt VectorXd U = - A.inverse() * B; }
\end{center}

For further details we refer the interested reader to the documentation of the source code on \href{http://doxygen.just-intonation.org}{\tt doxygen.just-intonation.org}.


\end{document}